\documentclass[conference]{IEEEtran}
\IEEEoverridecommandlockouts
\usepackage{cite}
\usepackage{amsmath,amssymb,amsfonts}
\usepackage{algorithmic}
\usepackage{graphicx}
\usepackage{textcomp}
\usepackage{xcolor}
\def\BibTeX{{\rm B\kern-.05em{\sc i\kern-.025em b}\kern-.08em
    T\kern-.1667em\lower.7ex\hbox{E}\kern-.125emX}}
\begin{document}

\onecolumn

\
\vfil
\noindent Preprint Notice:
\newline
© 2019 IEEE. Personal use of this material is permitted. Permission from IEEE
must be obtained for all other uses, in any current or future media, including
reprinting/republishing this material for advertising or promotional purposes,
creating new collective works, for resale or redistribution to servers or lists, or
reuse of any copyrighted component of this work in other works.
\vfil

\twocolumn
\newpage
\title{Weakly Supervised Fine Tuning Approach for Brain Tumor Segmentation Problem\\
}

\author{\IEEEauthorblockN{Sergey Pavlov}
\IEEEauthorblockA{\textit{Moscow Institute}\\
\textit{of Physics and Technology}\\
Moscow, Russia\\
sergey.pavlov@phystech.edu} \\

 \IEEEauthorblockN{Alexander Bernstein}
\IEEEauthorblockA{\textit{Skolkovo Institute}\\
\textit{of Science and Technology}\\
Moscow, Russia \\
a.bernstein@skoltech.ru}

\and 

\IEEEauthorblockN{Alexey Artemov}
\IEEEauthorblockA{\textit{Skolkovo Institute}\\
\textit{of Science and Technology}\\
Moscow, Russia \\
a.artemov@skoltech.ru} \\

\IEEEauthorblockN{Evgeny Burnaev}
\IEEEauthorblockA{\textit{Skolkovo Institute}\\
\textit{of Science and Technology}\\
Moscow, Russia \\
e.burnaev@skoltech.ru}

\and

\IEEEauthorblockN{Maksim Sharaev}
\IEEEauthorblockA{\textit{Skolkovo Institute}\\
\textit{of Science and Technology}\\
Moscow, Russia \\
m.sharaev@skoltech.ru} 

\thanks{}
}

\maketitle

\begin{abstract}
Segmentation of tumors in brain MRI images is a challenging task, where most recent methods demand large volumes of data with pixel-level annotations, which are generally costly to obtain. In contrast, image-level annotations, where only the presence of lesion is marked, are generally cheap, generated in far larger volumes compared to pixel-level labels, and contain less labeling noise. In the context of brain tumor segmentation, both pixel-level and image-level annotations are commonly available; thus, a natural question arises whether a segmentation procedure could take advantage of both. In the present work we: 1) propose a learning-based framework that allows simultaneous usage of both pixel- and image-level annotations in MRI images to learn a  segmentation model for brain tumor; 2) study the influence of comparative amounts of pixel- and image-level annotations on the quality of brain tumor segmentation;
3) compare our approach to the traditional fully-supervised approach and show that the performance of our method in terms of segmentation quality may be competitive.
\end{abstract}

\begin{IEEEkeywords}
MRI, Brain, Oncology/tumor, Image segmentation, Supervised learning, Weakly-supervised learning 
\end{IEEEkeywords}

% ely (\textit{fully}) 
% \textit{weakly supervised annotations}

\section{Introduction}
Gliomas represent approximately 30\% of all brain and central nervous system tumors and 80\% of all malignant brain tumors~\cite{genetics_blast}.
Glioblastoma, a grade IV glioma, is the most common and most aggressive primary brain tumor. It comprises 15\% of all intracranial neoplasms and 60-75\% of astrocytic tumors~\cite{trends}. 
Glioblastoma has the lowest 5-Year Relative Survival Rate compared to the other types of brain tumor: it varies from 19\% to 5\% for different 
age groups~\cite{survival_rates}. Once a glioma becomes symptomatic, patients' chance of recovery drops drastically. That's why early detection becomes crucial in the case of gliomas~\cite{gliomas_intro}.
Computer-assisted diagnosis (CAD) methods should reduce the cost and accelerate the process of lesion detection and segmentation. Recently
this problem has been widely studied, particularly thanks to BRATS challenges~\cite{brats_1,brats_2}, and their provided data sets. Despite that, the necessity of 
acquisition of large representative data sets rests unchanged. Pixel-wise delineation of lesions is a costly task
demanding significant resources. In this work, we propose a method that allows reducing the number of needed images with pixel-wise labels.

The article has the following structure. In Section \ref{related} we describe related works. Data and pre-processing are considered in Section \ref{sec:data}. We provide methods description in Section \ref{methods}. Resuts are  given in Section \ref{results} and conclusions are in Section \ref{conclusion}.

\section{Related Work}
\label{related}
Medical imagery segmentation task comprises several different problem statements depending on the type of the available in every particular case annotation of the training dataset. Here we introduce main options and give examples of works representing each of them.

\subsection{Fully-Supervised Learning}

The most natural and informative way to represent a lesion in a 2D image is to give it's precise contour. In the discrete case where images have a finite dimension and contain pixels, each pixel can be classified as belonging to the lesion or not. If this information is available for the whole dataset, this kind of annotation is called \emph{pixel-wise annotations} or \emph{full annotation}. In the particular case of brain MRI images, the challenge of BRATS uses this type of annotation. The winning solution \cite{myronenko_3d_2018} of BRATS-2018 uses encoder-decoder architecture with added variational autoencoder regularization. 

\subsection{Weakly-Supervised Learning}
When the full annotation is unavailable we may still extract some useful information for segmentation from what is called \emph{weak annotations}. For example, bounding box annotation \cite{rajchl_deepcut:_2016} is a type of weakly supervised annotation. More commonly the image-level annotations are used (presence/absence of lesions for example). Multiple Instance Learning (MIL) method\cite{pathak_fully_2014} has recently gained much attention\cite{quellec_multiple-instance_2017} and has become one of classical approaches to train segmentation with weak supervision\cite{zhu_deep_2016}. MIL can be generalized \cite{kervadec_constrained-cnn_2019} in the sense that it also introduces geometrical constraints on pixel map. Another way to incorporate geometric properties of lesions is based on CNN detector (i.e. binary classification) architectures \cite{rongchang_weakly-supervised_2018-1,maicas_model_2018}. 
\newline
\indent The MIL method consists of four principal aspects:
\begin{itemize}
    \item A feature map $\mathcal{M} \in [0,1]^{m\times n}$ is somehow obtained from the initial image (e.g. with the techniques of CNN); these values are interpreted as probabilities of presence of lesion in a particular pixel of the feature map;
    \item A positive value ${\bf K}\in \mathbb{N}$ is chosen. Often some knowledge of the problem structure (e.g. average size of lesions in pixels of feature map) may be useful for this choice;
    \item Values of feature map $\mathcal{M}$ are sorted in descending order: $a_1 \ge a_2 \ge \dots \ge a_{mn}$;
    \item A specific {\bf loss function} to optimize is constructed in the following way: if the whole image is classified positively (contains lesion), it is the sum of pixel-wise log-losses over the pixels of the feature map $\mathcal{M}$ where first {\bf K} pixels (i.e. corresponding to $a_1$ to $a_{\bf K}$) are considered positively classified (so 1 is used as true value in the corresponding log-loss) and resting $mn - {\bf K}$ pixels are considered negatively classified (so 0 is used as true value in the corresponding log-loss). If the whole image is classified negatively (contains no lesion) all log-losses are calculated with respect to the true value of 0. 
\end{itemize}
In the present work $m=4, n=3$, so the feature map $\mathcal{M}$ contains $4\times3 = 12$ pixels and $ {\bf K}=4$. The details of choice of parameter {\bf K} as well as a more formal description of the MIL method will be provided in the section \ref{methods}.

\subsection{Semi-Supervised Learning}
If fully annotated images are still available but their amount is relatively small with respect to the size of training dataset, we could attempt to extract some additional information (for example for prediction of size of the target region) from the part of the dataset not containing annotations\cite{kervadec_curriculum_2019}. This type of learning is called \emph{semi-supervised}.

The whole dataset could also be weakly annotated along with provided full annotation for some instances. There is an attempt to solve this problem with GANs\cite{vorontsov_boosting_2019}. We will use a more straightforward way by introducing an architecture that allows to use the fully supervised learning with MIL method training.

\section{Data and Preprocessing}
\label{sec:data}

We work with the data of the BRATS-2018 competition \cite{BRATS_2018}, which contains MRI images for 285 patients diagnosed with glioma. For each patient, several series of slices represent different projections and contrast of MRI scanning. Each slice is annotated with a per-pixel segmentation mask. Although BRATS-2018 contains segmentation masks for various tumor parts, we target the problem of {\bf whole tumor} segmentation only. In this work, the post-contrast T1-weighted scan (T1Gd) images were used. We preprocess the selected slices as follows: 
\begin{itemize}
    \item[1)] we omit near-boundary slices that contain few bright pixels; 
    \item[2)] we align each series by most-right, -left, -upper, and -down achieved borders to reduce the area of background pixels; 
    \item[3)] we resize each series to 170x140 resolution. 
\end{itemize}
The number of images per patient varies between 98 and 123, with an average of 110. The percentage of images containing lesions varies from 18\% to 90\%, with an average of 56\%. Overall, there are 31350 images, where 15761 contain lesions.

We introduce three groups of images, according to the minimal lesion percentage area: ``$\geqslant 1\%$'', `` $\geqslant 5\% $'', ``$\geqslant 10\%$'' limiting lesions (only for ``positive'' images, i.e. containing lesions) make up at least respectively 1\%, 5\% or 10\% of the total brain cross-section area of the image, so that the number of ``positive'' images in the groups sums up to respectively 13806, 9804 and 5636 correspondingly. In addition to selected ``positive'' images, all ``negative'' images (i.e. images not containing lesions) are also included into each of these groups.

\section{Methods}
\label{methods}
\subsection{Architecture Description}
\label{sec:architecutre} 

\begin{figure}[htbp]

\centerline{\includegraphics[scale=0.35]{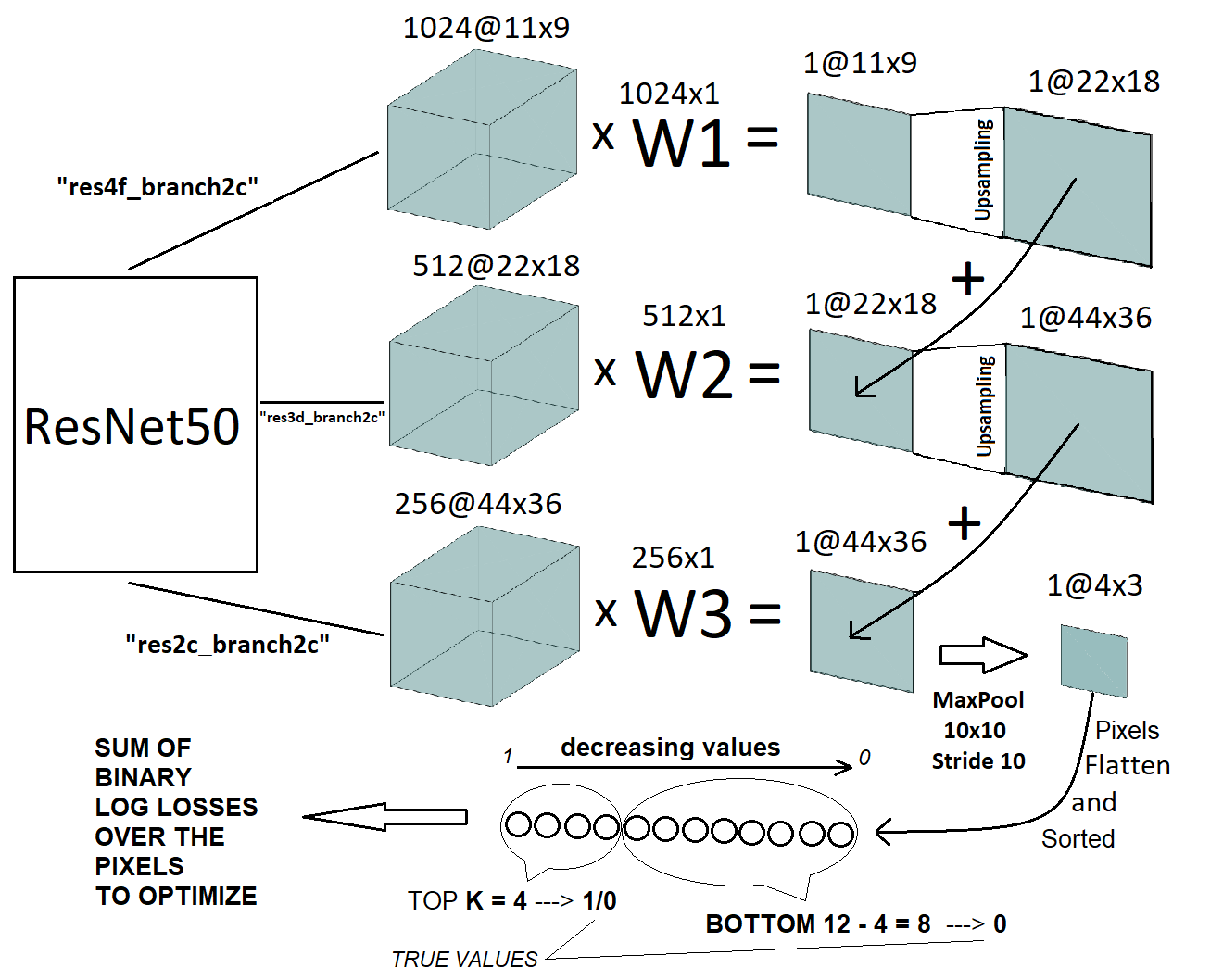}}

\caption{The architecture structure is organized in three steps. Each step consists of: 1) Weighted integration; 2) Upsampling. The output of each step is added to the output of the following step. }
\label{fig:schema}
\end{figure}

Our architecture is inspired from \cite{rongchang_weakly-supervised_2018-1} and  based on ResNet-50\cite{he_deep_2015} network (Keras Applications implementation). We take feature maps $M^1\in \mathbb{R}^{1024\times11\times9}$, $ M^2\in \mathbb{R}^{512\times22\times18}$, $ M^3 \in \mathbb{R}^{256\times 44\times36}$ of layers ``res4f\_branch2c'', ``res3d\_branch2c'' and ``res2c\_branch2c'' of ResNet-50 respectively and sum them up correspondingly with trainable weights $W^1\in\mathbb{R}^{1024\times1}, W^2\in\mathbb{R}^{512\times1}, W^3\in\mathbb{R}^{256\times1}$ step-wise. At each step $i$ a map $A^i$ is defined as a sum of the output of the upsampled (by a factor of 2 in both axes) output $A^{i-1}$ of the previous step and a weighted summation of feature maps corresponding to this step. We also apply valid max-pooling with filter size $10\times10$ and stride 10 to the output $A^{3}$ of the last step and introduce $\tilde{A}^3$.  This technique allows to extract more abstract features from deeper layers along with less abstract features from less deep ones. Formally the architecture (Fig.~ \ref{fig:schema}) can be written as follows:

$$A^1 := \sum_{i=1}^{1024}W^1_i M^1_{i,\cdot,\cdot}\in \mathbb{R}^{11\times9},$$

$$A^2 := \sum_{i=1}^{512}W^2_i M^2_{i,\cdot,\cdot} + UpSampling(2,2)[A^1]\in \mathbb{R}^{22\times18},$$

\begin{align*}
    A^3 := \sigma\left(\sum_{i=1}^{256}W^3_i M^3_{i,\cdot,\cdot} + UpSampling(2,2)[A^2]\right)\\
    \in [0,1]^{44\times36},
\end{align*}

$$\tilde{A}^3 := MaxPool(10\times10, stride = 10)[A^3] \in [0,1]^{4\times3},$$
where~ $\sigma(\cdot)$ is a sigmoid function.

\subsection{Method Description}
\label{sec:method}

In what follows, we call learning with pixel-level annotations \emph{fully-supervised} (FS) and its corresponding annotations \emph{full}, while learning with image-level annotations is called \emph{weak-supervised} (WS), and its respective annotations \emph{weak}.

\begin{figure}[htbp]
\centerline{\includegraphics[scale=0.40]{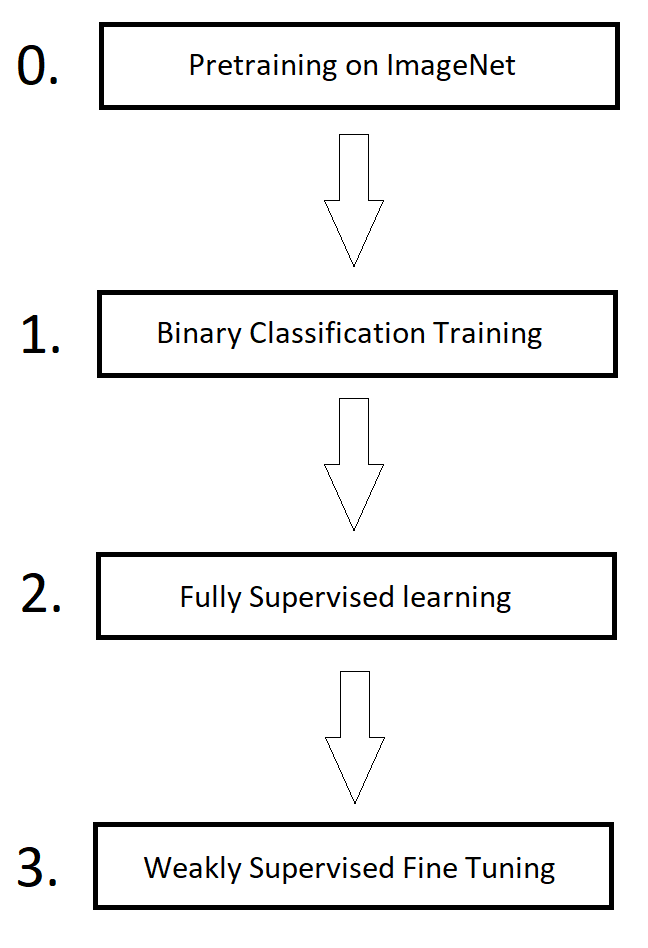}}
\caption{Approach steps}
\label{fig:schema_eng}
\end{figure}

\begin{figure}[htbp]
\centerline{\includegraphics[scale=0.50]{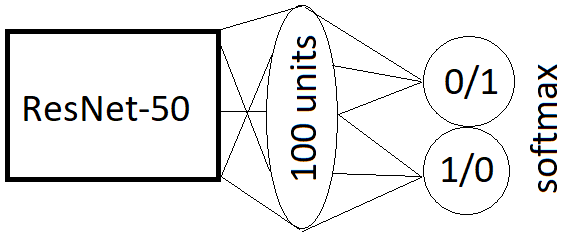}}
\caption{Classification architecture}
\label{fig:class_arch}
\end{figure}

As shown in the Fig.~ \ref{fig:schema_eng}, our method consists of two auxiliary steps and two main steps. The auxiliary steps are:

\begin{itemize}
    \item {\bf Pretraining on ImageNet}: we initialize the underlying ResNet-50 weights with weights pretrained on ImageNet dataset;
    \item {\bf Binary classification pretraining step}: using only weak supervision, we may pretrain the original ResNet-50 on the classification task using the architecture shown in the Fig.~\ref{fig:class_arch}. This step could familiarize our network with the features of the new dataset and facilitate the training on the main steps. More details on this pretraining are given in Section~\ref{sec:bin_cl};
\end{itemize}

Here we describe the two main steps.

\subsubsection{Fully Supervised Learning Step}
Fully supervised segmentation is traditionally learned with Dice index:
$$\text{Dice~index}(P,T) := \frac{2\sum_{i,j}P_{ij}T_{ij} + 1}{\sum_{i,j}P^2_{i,j} + \sum_{i,j}T^2_{i,j} + 1},$$
where $P$ is the pixel map prediction ($P_{i,j}\in [0,1]$), $T$ is a ground truth binary pixel map ($T_{i,j}\in\{0,1\}$). We define also Dice loss function:

$$\text{Dice~loss}(P,T) := 1 - \text{Dice~index}(P,T)$$

In our case: $P \equiv A^3$, so we train our segmentation on $44\times36$ feature maps, downsampled from the original shape of $170\times140$ introduced in Section~\ref{sec:data}.

Fully supervised step will be performed on a possibly small datasets using data augmentation (rotations, translations). The dependency of the final result on the size of those datasets will be studied in Section~\ref{sec:results}.

\subsubsection{Weakly Supervised Learning Step}

The weakly supervised learning step is based on the Multiple-Instance learning (MIL)\cite{pathak_fully_2014} and performed on $\tilde{A}^3$ as shown in the Fig.~\ref{fig:schema}. Let $(a_1, a_2,\dots, a_{12}) := sort[(\tilde{A}^3_{i,j})_{1\leqslant i \leqslant 4, 1\leqslant j \leqslant 3}]$ --- sort in descending order. Let $I_n$ define $n^{th}$($1 \leqslant n \leqslant N$) training image, $y_n \in\{0,1\}$ --- it's label (absence/presence of lesion), $I_n^i$ --- the part of $I_n$ on which $a_i$ depends, and $\theta$ --- trainable parameters of the architecture including $W^1, W^2$ and $W^3$. So we define:
$$p(y=1|I_n^i, \theta):= a_i, ~p(y=0|I_n^i, \theta):= 1 - a_i.$$

The MIL method is performed on $\tilde{A}^3$ which consists of $4\times3 = 12$ pixels. Prior to training parameter $K\in \mathbb{N}$ is chosen. If a particular image contains lesions (i.e. a global label $y=1$ is attributed to it), a label of 1 is used as a ``true'' values to exactly $K$ top valued pixels. Other $12 - K$ pixels are attributed with a ``true'' value of 0. In the case when $y=0$, all 12 pixels of $\tilde{A}^3$ are attributed with a ``true'' value of 0. The defined   ``true'' value of each pixel is next used to calculate a sum of binary cross-entropy losses of different pixels. Here we describe the MIL loss formally:

\begin{equation*}
  \label{eq:equation1}
  \mathcal{L} := -\frac{1}{12\times N}\sum_{n=1}^N\left( \sum_{j=1}^K\log(p(y_n|I^{j}_n,\theta)) + \right.
\end{equation*}
\begin{align*}
  +\left.\sum_{j=K+1}^{12}\log(p(y=0|I^{j}_n,\theta))\right)
\end{align*}
In addition to $\mathcal{L}$ we consider a l2-regularization ($\lambda = 5e^{-6}$) term over the kernel (not biases) parameters of ResNet-50.
We conducted a series of 5-fold cross validation experiments for different values of parameter $K$ and we chose the values of  $\bf K=4$ as the optimal value in terms of Dice loss.

\subsection{Training Details}
The learning was conducted in the following way:
\begin{itemize}
    \item {\bf Binary classification} step is trained with Early Stopping 3 epochs (i.e. we stop training when no improvement of validation loss is observed during 3 consecutive epochs);
    \item  {\bf Fully supervised} learning step is trained with Early Stopping 15 epochs; then we restore best weights and launch a new convergence with 10 times smaller learning rate and  with Early Stopping 5 epochs;
    \item  {\bf Weakly supervised} learning step is trained with Early Stopping 2 epochs and a tolerance value of 0.01; If there is no improvement from the very beginning we restart training with a 10 times smaller learning rate;
\end{itemize}

We used Adam optimizer\cite{kingma_adam:_2014} for training. In binary classification step we used $\alpha^{bin} := 5\times10^{-5}$. Learning rates  $\alpha^{fs}$ and $\alpha^{ws}$ were selected empirically based on observations of the convergence behaviour as a function of the number $N_{fs}$ of images used for the fully supervised learning step using the following heuristic:

$$
\alpha^{fs} = \frac{5\times10^{-2}}{N_{fs}},
\qquad
\alpha^{ws} = \frac{5\times10^{-3}}{N_{fs}^2}.
$$
The performance of FS step significantly depends on the choice of $\alpha^{fs}$. One may observe in Fig.\ref{fig:main_result_10}, \ref{fig:main_result_5}, \ref{fig:main_result_1} that the behaviour of the FS curve (distance between curves) varies between different datasets, whereas the WS step behaves more robustly.
For the reversed order (WS step followed by FS step) of training (discussed in Section~\ref{sec:Order})
we used $\alpha^{fs}:= 5\times10^{-6}$ and $$\alpha^{ws}:= \frac{5\times10^{-6}}{N_{fs}}.$$

For the most of experiments 5-fold cross validation was applied. At each cross validation step 228 patients were used for training and other 57 for testing. For WS step both training and test sets were balanced so that they contain doubled number of positively (negatively) diagnosed images. For FS step a certain number $N_{fs}$ of WS were taken with their pixel-wise segmentation as well as exactly $N_{fs}$ images diagnosed negatively. For the rest of the present work this will be noted as ``{\bf number of images}, ${\bf \times2}$'' or ``$\bf N_{fs}(+N_{fs})$''.

\section{Results}
\label{results}
\subsection{Binary Classification Pretraining Performance}\label{sec:bin_cl}

\begin{figure}[htbp]
\centerline{\includegraphics[scale=0.5]{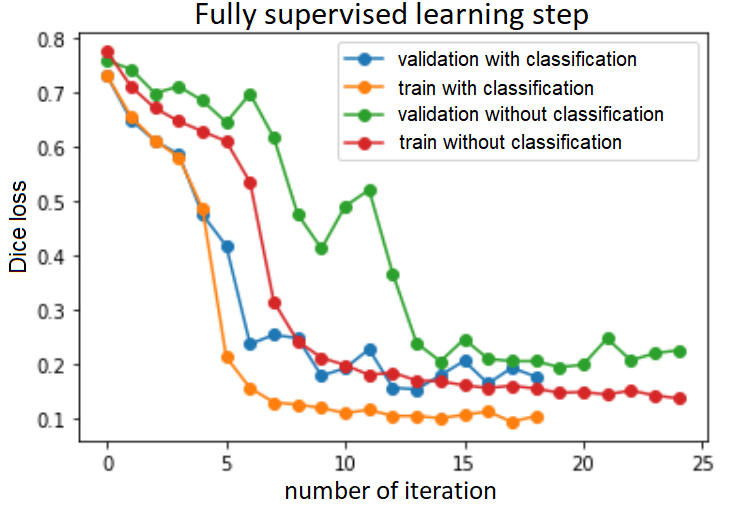}}
\caption{Learning curves with binary classification pretraining and without}
\label{fig:cmp_class_no_class}
\end{figure}

As one can see in Fig.~\ref{fig:cmp_class_no_class}, the binary classification pretraining provides faster convergence at the fully supervised learning step and leads to better results.  

\begin{figure}[htbp]
\centerline{\includegraphics[scale=0.5]{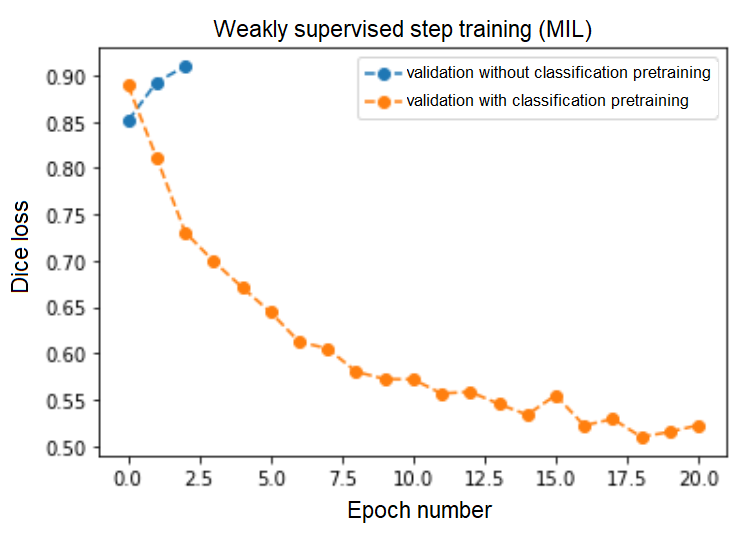}}
\caption{Learning curves for weakly supervised learning step when WS step precedes FS step}
\label{fig:cmp_order_mil}
\end{figure}

Moreover (see Fig.~\ref{fig:cmp_order_mil}) the binary classification pretraining is crucial for convergence of weakly supervised step when steps are reversed (WS step precedes FS step).

\subsection{Order Of Steps}
\label{sec:Order}

\begin{figure}[htbp]
\centerline{\includegraphics[scale=0.5]{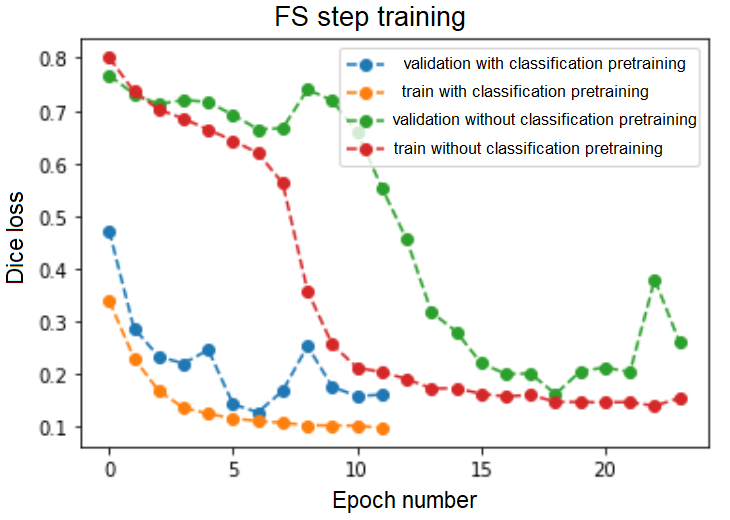}}
\caption{Learning curves for fully supervised learning step when FS step follows WS step}
\label{fig:cmp_order_class}
\end{figure}

We compare performance of the two main steps both in cases with binary classification pretraining of ResNet-50 and without (for the ''10\%'' dataset, 500(+500) setup). 
In Fig.~\ref{fig:cmp_order_class} and Table \ref{tbl3} one can observe that the order chosen above is indeed optimal as well as the binary classification pretraining step utilization. 
Learning curves for the reversed case (i.e. MIL step is trained before FS step) are shown in Fig.~\ref{fig:cmp_order_mil} and Fig.~\ref{fig:cmp_order_class}. With binary classification pretraining step FS step converges much faster and without the WS step does not converge at all.

\begin{table}[htbp]
\caption{ Steps order comparison}
\scriptsize
\begin{center}
\resizebox{\columnwidth}{!}{%
\begin{tabular}{|c|c|c|c|} 
 \hline
  Order of steps & Classification pretraing  & Dice loss, step 1 & Dice loss, step 2 \\
 \hline
 FS $\rightarrow$ MIL & YES & $0.15 \pm 0.02$ & $0.14 \pm 0.02$ \\ 
 \hline
 FS $\rightarrow$ MIL & NO & $0.20 \pm 0.04$ & $0.17 \pm 0.02$ \\
 \hline
 MIL $\rightarrow$ FS & YES & $0.47 \pm 0.19$ & $0.15 \pm 0.03$ \\ 
 \hline
 MIL $\rightarrow$ FS & NO & $0.84 \pm 0.03$ &  $0.19 \pm 0.03$ \\ 

 \hline

\end{tabular}
}
\end{center}
\label{tbl3}
\end{table}

\subsection{Principal Results}\label{sec:results}

\begin{figure}[htbp]
\centerline{\includegraphics[scale=0.5]{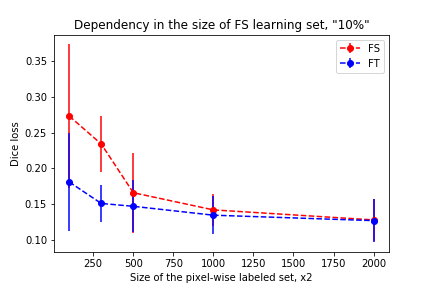}}
\caption{Results for fully supervised learning  step (red) and fine tuning MIL step (blue) on dataset ''10\%'' with 95\%-confidence intervals;}
\label{fig:main_result_10}
\end{figure}

\begin{figure}[htbp]
\centerline{\includegraphics[scale=0.5]{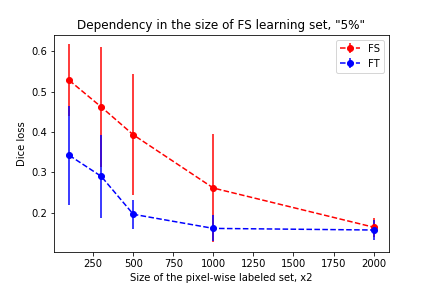}}
\caption{Results for fully supervised learning  step (red) and fine tuning MIL step (blue) on dataset ''5\%'' with 95\%-confidence intervals;}
\label{fig:main_result_5}
\end{figure}

\begin{figure}[htbp]
\centerline{\includegraphics[scale=0.5]{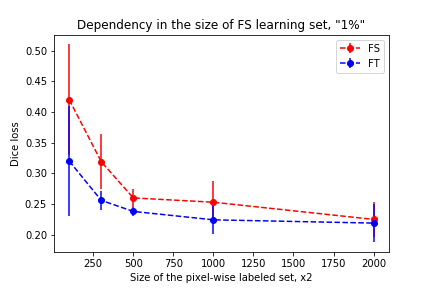}}
\caption{Results for fully supervised learning  step (red) and fine tuning MIL step (blue) on dataset ''1\%'' with 95\%-confidence intervals;}
\label{fig:main_result_1}
\end{figure}

We report main results of this work.
Firstly, we compare the straight order (i.e. MIL after FS) algorithm  results for different sizes of pixel-wisely labeled sets used at the FS step (see Fig.~\ref{fig:main_result_10}, \ref{fig:main_result_5}, \ref{fig:main_result_1}). We remind that we are using balanced training datasets for the FS step and it's size is twice the size of corresponding pixel-wise labelled set containing tumor (the second half corresponds to clear images not containing tumor).

Secondly, we compare our method (500($250+250$) fully annotated images) to purely fully supervised approach (see Table \ref{tbl2}). The results are provided for ``$\ge1\%$’’, ``$\ge5\% $’’ and ``$\ge10\% $’’ datasets and for the 80\% (training)/20\% (test) splits for the purely fully supervised approach.

Our approach allows to reduce the number of pixel-wisely segmented images containing tumor up to 150-250, still providing competitive quality of segmentation. Moreover, it demonstrates robustness with respect to a possible non-optimality of chosen learning rate values for different data sets.

The obtained quality of segmentation is not uniform over different datasets. Adding slices with smaller tumor cross section area makes the problem more difficult. 

\begin{figure}[t!]
\centerline{\includegraphics[scale=0.4]{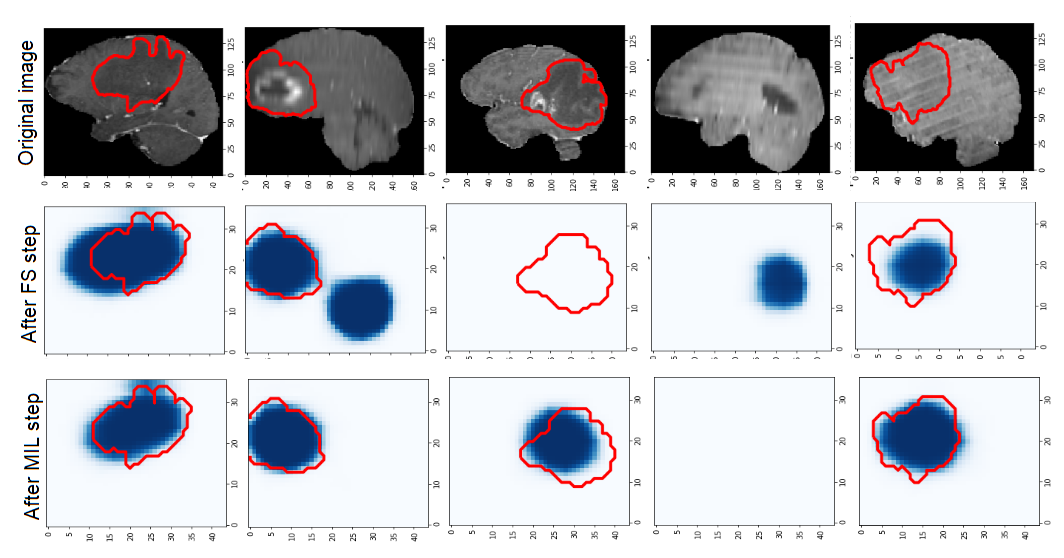}}
\caption{Examples of errors of FS step corrected by MIL step. First row: original images with ground truth {\bf original segmentation} (red contour) where available; Second row: segmentation (blue) with downsampled ground truth contour (red) after {\bf FS step}; Trird row: segmentation (blue) with downsampled ground truth contour (red) after {\bf MIL step};  }
\label{fig:cmp_fs_our}
\end{figure}

We also report the visual analysis results of comparison of two steps of our approach: FS and FT. Among 100 randomly selected validation set images we found 22 cases of improvement of the whole tumor segmentation quality after the fine tuning (FT) step compared to the preceding FS step, 3 cases showed changes in the opposite direction and for other cases the quality seems to be the same. Some examples of segmentation errors corrected between steps are shown in Fig. \ref{fig:cmp_fs_our}.

\begin{table}[htbp]
\caption{Our method (FS+MIL)  for 250 fully annotated images(+250 without lesions) compared to the purely fully supervised approach on the whole dataset}
\begin{center}
\begin{tabular}{|c|c|c|} 
 \hline
  Dataset & FS+MIL, 5-fold CV  & Fully supervised, $80\%/20\%$\\
 \hline
 $\geqslant 1\%$ & $0.24\pm0.01$ & 0.2099  \\ 
 \hline
 $\geqslant 5\%$ & $0.2\pm0.03$ & 0.1629 \\
 \hline
 $\geqslant 10\%$ & $0.15\pm0.04$ & 0.1077  \\ 
 \hline

 \hline

\end{tabular}

\end{center}

\label{tbl2}

\end{table}

\section{Conclusion}
\label{conclusion}

We proposed a framework that allows to use both weak and pixel-wise annotations to learn segmentation. We also demonstrated that our method may reduce the need of costly fully annotated images by using the weakly annotated ones. 

As a possible direction for further research we view alternating between fully- and weakly- supervised steps during training, which should provide a gradual increase in segmentation quality. Another possible direction could be an improvement the MIL feature map acquisition method (instead of the used MaxPooling technique). 

We can also use a special boundary loss from \cite{boundary} and a fusion of multi-fidelity data \cite{Multispectral2018} to increase semantic segmentation accuracy; sparse convolutions from \cite{3DCNN2018} to increase computational efficiency. 
Another possibility is to apply bayesian generative models from \cite{kuzina} and latent convolutional manifolds from \cite{latent} for transfer learning of semantic segmentation tasks.

A challenging task would be to test the developed capabilities on other types of image data, e.g. from remote sensing applications \cite{procedural,RSdamage2018}.

The study was supported by the Russian Science Foundation under Grant 19-41-04109.

%We consider the additional study to be of great interest both for segmentation accuracy improvement and reduction of the need in labeled data.

\bibliographystyle{./bibliography/IEEEtran}
\bibliography{./bibliography/IEEEabrv,./bibliography/IEEEexample}

\end{document}